# Designing wearable haptic cueing for neurological gait rehabilitation: user requirements across residential care and community settings

Riasat Islam — Queen Mary University of London, London, UK
Annemarie Zijlema — University of Greenwich, London, UK
Theodoros Georgiou — Heriot-Watt University, Edinburgh, UK
Simon Holland — The Open University, Milton Keynes, UK
Blaine Price — The Open University, Milton Keynes, UK
Paul Mulholland — The Open University, Milton Keynes, UK
Oliver Pearce — Milton Keynes University Hospital, Milton Keynes, UK
Allan Perry — PJ Care, Milton Keynes, UK
Corresponding author: Riasat Islam; riasat.islam@qmul.ac.uk

Word count: 9,038 for Introduction through Conclusion, including headings, tables and figure captions; 11,418 from Abstract through References, including declarations and keywords. Abstract: 250 words including headings (244 excluding headings).

## Abstract

**Purpose:** We identified user, design, and implementation requirements for wearable haptic cueing in neurological gait rehabilitation. Haptic cues leave hearing available for communication and environmental awareness, but their use depends on more than cue delivery.

**Materials and Methods:** We analysed six linked studies involving 17 people with neurological conditions affecting gait and three physiotherapists across formative sessions, residential care, and community self-management. Contrasting wearable systems were used as technology probes to expose requirements through differences in haptic output, form factor, attachment, and interaction. Observation notes, interviews, diaries, physiotherapist records, gait analysis, and activity logs informed thematic analysis and cross-study comparison.

**Results:** Seven recurring requirement areas concerned haptic perceptibility and placement, wearability and fastening, interaction burden, rhythm personalisation, real-world walking contexts, supported use, and perceived value beyond gait outcomes. Requirements interacted: locations where cues were most perceptible could be difficult to access for fastening or control. Fixed rhythms did not always accommodate terrain, fatigue, interruptions, or changing pace. Carers and clinicians supported configuration, operation, safety, and interpretation. Participants described confidence, routine formation, and willingness to walk; these accounts were interpreted as perceived value rather than evidence of therapeutic efficacy. Boundary cases also showed that low sensation, cognitive burden, and mobility limitations could make wearable cueing impractical.

**Conclusion:** Future systems should combine adjustable haptic output, flexible placement, accessible fastening, simple controls, personalised rhythms, and robustness during everyday walking. Design should accommodate support from carers and clinicians and consider trade-offs between requirements. These findings inform rehabilitation-oriented wearable design rather than establish clinical efficacy.

## Implications for rehabilitation

- Wearable haptic cueing systems should provide adjustable haptic output and flexible body placement, because users may differ in where and how reliably they can perceive rhythmic cues.
- Rehabilitation-oriented haptic devices should use accessible fastening, low-burden controls, and clear device feedback so that people with sensory, dexterity, or cognitive limitations can use them safely and consistently.
- Rhythm settings should be personalised and adaptable to real-world walking conditions, including fatigue, route and terrain changes, interruptions, and changes in walking pace.
- Carers and clinicians should be considered as a central part of the use process, with clear workflows for setup, troubleshooting, safety monitoring, and interpretation of use.

## Introduction

Gait impairment is common across neurological conditions such as stroke, acquired brain injury, Parkinson's disease, multiple sclerosis, and Huntington's disease. Changes in gait speed, symmetry, foot clearance, balance, and endurance can limit community mobility, increase fall risk, and reduce independence and quality of life (Belda-Lois et al., 2011; Baker, 2018; Bilney et al., 2003; Hollands et al., 2012; Winstein et al., 2016). These difficulties also matter beyond individual mobility because falls, reduced independence, and reduced participation can increase the need for rehabilitation, supervision, informal care, and health and social care support.

Rhythmic cueing has long been used in rehabilitation to support movement timing and coordination. Cueing can use different sensory modalities, including auditory, visual, and somatosensory cues. Auditory cueing, often delivered through metronomes or music, has received substantial attention in gait rehabilitation (Thaut et al., 1997; Conklyn et al., 2010; Hove et al., 2012; Ghai et al., 2018). Wearable haptic cueing offers a different route because it can deliver rhythmic prompts through vibration on the body while leaving the auditory channel free for conversation, environmental awareness, and safety (Demain et al., 2013; Holland et al., 2014; Georgiou et al., 2015; Chen et al., 2017).

However, wearable haptic cueing must work as more than a cue generator if it is to support rehabilitation practice. Users need to perceive the cue reliably, wear the device at a suitable body location, fasten and operate it independently, personalise or follow the rhythm, and use it safely in real walking contexts. These requirements become especially important in neurological rehabilitation, where users may have sensory loss, hemiparesis, reduced dexterity, cognitive burden, fatigue, balance concerns, or changing walking patterns.

Existing reviews of wearable haptic cueing show growing clinical interest, but the field has focused mainly on feasibility, technical performance, and gait or motor outcomes. Reviews and meta-analyses have primarily synthesised effects on gait, balance, freezing of gait, and motor function, especially in Parkinson’s disease and other neurological conditions (Sweeney et al., 2019; De Angelis et al., 2021; Gonçalves et al., 2021; Zhang et al., 2024). Earlier review and feasibility literature has similarly focused on vibrotactile cue delivery, freezing-of-gait management, or short-term motor effects rather than on what users need to wear, configure, and sustain cueing in practice (Casamassima et al., 2014; Gonçalves et al., 2017; Li et al., 2023).

This gap matters because a wearable haptic cueing system that appears promising in a supervised or controlled setting may still fail in practice if it assumes intact sensation, easy fastening, reliable gesture control, stable walking cadence, or independent use. Wearable rehabilitation technologies also need to be interpretable, supportable, and usable in everyday settings rather than only in controlled assessments (Brodie et al., 2016; Iosa et al., 2016; Díaz et al., 2019; Keogh et al., 2021). We therefore need research that uses wearable haptic devices not only to test cueing effects, but also to understand the design space for future rehabilitation-oriented systems.

This paper analyses six linked studies of wearable haptic cueing across three rehabilitation contexts: formative user sessions, residential-care rehabilitation, and community self-management. Across these settings, we examined how people with neurological gait impairment perceived, wore, controlled, configured, and integrated wearable haptic rhythms into walking practice. The main contribution is therefore not only the identification of individual requirements, but showing how those requirements interact across real rehabilitation contexts.

We used repurposed off-the-shelf wearable haptic devices as technology probes. Technology probes support design research by introducing exploratory technologies into real-world settings to understand user needs, collect use data, and inspire future design (Hutchinson et al., 2003; Boehner et al., 2007). We did not treat the probe devices as finished rehabilitation interventions. Instead, we used them to identify user, device, and implementation requirements for future wearable haptic cueing systems.

This paper builds on earlier wearable haptic cueing work in neurological rehabilitation, including prototype development, formative studies, and case-based deployments (Islam et al., 2018; Georgiou et al., 2020a; Georgiou et al., 2020b; Islam, 2023). However, rather than reporting a single prototype or case, this paper blends cross-setting requirements for perception, placement, fastening, interaction, rhythm configuration, real-world walking, and supported use.

This paper addresses two research questions:

1. What recurring user barriers emerged when wearable haptic cueing devices were used across formative user sessions, residential-care rehabilitation, and community self-management?
2. What user, device, and implementation requirements can be derived for future wearable haptic cueing systems in neurological gait rehabilitation?

# Materials and methods

## Study design

We conducted six linked studies of wearable haptic cueing for neurological gait rehabilitation. The studies differed in setting, duration, participant needs, level of supervision, device configuration, and available data sources. We designed this variation deliberately. We did not aim to test one device, one neurological condition, or one clinical setting. Instead, we aimed to identify user, device, and implementation requirements across a broad range of neurological rehabilitation contexts.

We began with formative user sessions involving participants with different neurological conditions. These sessions allowed us to explore early issues around haptic perception, body placement, fastening, interaction, and rhythm use across a heterogeneous group. We then moved into a residential-care setting, where physiotherapists supported the use of wearable haptic cueing with residents who had more complex mobility, sensory, and cognitive needs. Finally, we examined community self-management, where participants used wearable haptic cueing in everyday walking routines with different levels of carer, clinician, or researcher support.

This staged design followed a pragmatic, design-oriented approach to HCI research (Fallman, 2003; Holtzblatt & Beyer, 1997). We used the studies to understand how wearable haptic cueing worked in practice, what barriers emerged across settings, and what future rehabilitation-oriented systems would need to support. We did not design the studies as efficacy trials. Instead, we used them to identify recurring requirements and important boundary conditions.

As outlined above, we used wearable haptic devices as technology probes rather than finished rehabilitation interventions. In terms of methodology, this meant treating the Apple Watch, Soundbrenner Pulse, and Peterson BodyBeat Pulse Solo as contrasting probes for perception, placement, fastening, interaction, rhythm configuration, and support.

## Studies, participants, and data sources

The analysis included six linked studies across three rehabilitation contexts:

1. formative user sessions;
2. residential-care rehabilitation; and
3. community self-management.

All six studies examined the same core problem: how people with neurological gait impairment could perceive, wear, operate, configure, and integrate wearable haptic cueing into walking practice. The studies varied the level of supervision, user independence, environmental complexity, and support available. This variation helped us compare requirements that appeared across settings and identify barriers that emerged only under particular use conditions.

Across the six studies, 17 participants with neurological conditions and 3 physiotherapists contributed data. The participant group was heterogeneous by design. We included people with different neurological conditions because the aim was not to produce condition-specific efficacy evidence, but to identify user requirements for wearable haptic cueing across varied neurological rehabilitation contexts.

We report functional and contextual characteristics rather than full demographic details because several studies involved small samples and identifiable clinical contexts. Table 1 therefore focuses on the characteristics most relevant to transferability: participant profile, setting, device, duration, design-relevant characteristics, and data sources. We use pseudonyms for the four individual community participants. We report the formative and residential-care studies at group level because those studies generated findings across multiple participants within shared settings.

Table 1 summarises the included studies, rehabilitation contexts, and data sources.

*Table 1: Characteristics of the six linked wearable haptic cueing studies included in the cross-study analysis (names altered to protect the participants' anonymity).*

| Study | Participants and functional profile | Setting, device, and duration | Key design-relevant characteristics | Data sources |
|---|---|---|---|---|
| 1. Formative user sessions | 7 participants with brain injury, stroke, Parkinson's disease, multiple sclerosis, and one undisclosed neurological condition; mixed mobility profiles | Institutional and community sessions; Apple Watch with bespoke haptic metronome app and Soundbrenner Pulse; single sessions | Hemiparesis, reduced upper-limb dexterity, reduced haptic perception on the affected side, and mostly researcher-supported use | Observation notes; interview summaries |
| 2. Residential-care technology probe | 6 residents and 3 physiotherapists; mixed mobility, including supervised walking | Specialist neurological residential care; Soundbrenner Pulse; 2 weeks | Low residual haptic perception in some residents, limited dexterity, cognitive burden, and high physiotherapist involvement | Physiotherapist observation records; physiotherapist records; physiotherapist focus group |
| 3. Community multiple sclerosis study | Hannah, a participant with multiple sclerosis; used a walking stick outdoors and furniture support indoors | Community self-management; Soundbrenner Pulse; 3 weeks | Side-specific sensory difference, variable day-to-day mobility, fatigue, and self-managed use in a family context | Observation notes; interview transcript and summary; participant diary; limited walking/activity notes |
| 4. Community Huntington's disease study | Rosie, a participant with Huntington's disease, with support from her mother and friend | Community self-management; Peterson BodyBeat Pulse Solo; 3 weeks | Difficulty with watch-like attachment and controls; carer-assisted use | Observation notes; interview transcript; carer-supported interview material; participant diary; diary-based walking records |
| 5. Community stroke study | Tracy, a participant with stroke; community ambulator in hilly routes | Community self-management with gait-lab support; Soundbrenner Pulse; 2 weeks | Tempo mismatch between fixed cue rhythm and real walking route was more prominent than fastening difficulty; mostly self-managed use with research and physiotherapy input | Observation notes; interview transcript and summary; participant diary; physiotherapist comments; gait analysis; activity summaries |

| Study | Participants and functional profile | Setting, device, and duration | Key design-relevant characteristics | Data sources |
|---|---|---|---|---|
| 6. Community brain injury study | Bob, a participant with brain injury; independent community ambulator with residual gait impairment | Community self-management with gait-lab support; Soundbrenner Pulse; 2 weeks plus a repeat 2-week phase after 1 year | Reduced strength and dexterity in both arms, preference for stronger haptic output, clinician interpretation, and device modification support | Observation notes; interview transcript and summary; participant diary; physiotherapist or clinician comments; gait analysis; activity summaries |

Data sources varied between studies for methodological reasons. This reflected the pragmatic design of the research and the constraints of working across different rehabilitation contexts. The formative user sessions generated observation notes and interview summaries focused on perception, placement, fastening, and interaction. The residential-care technology probe generated physiotherapist records and focus-group data about suitability, supervised use, cognitive burden, and clinical workflow. The community self-management studies generated interviews, diaries, observation notes, and, where feasible, gait or activity data.

We treated each data source according to its strengths and advantages. Verbatim interview and focus-group transcripts supported direct participant or clinician accounts. Observation notes, diary summaries, and interview summaries supported the identification of design-relevant events, usability barriers, and contextual constraints. Gait analysis and activity data helped contextualise the qualitative findings, especially around mismatch between fixed cue tempo and variable real-world walking cadence, walking routines, and differences between laboratory and community walking.

The community studies (3–6 in Table 1) primarily contributed to understanding self-managed wearability, rhythm use, routine formation, and perceived value. The residential-care study (Study 2) contributed to understanding suitability, supervised use, clinician-mediated configuration, and conditions under which cueing was unsuitable. We retained themes when the available evidence supported cross-study interpretation or when a study exposed an important boundary condition for future design.

## Devices used as technology probes

We used three wearable systems as technology probes:

- Apple Watch Series 2 (Figure 1A) running a bespoke haptic metronome application (Apple Watch Series 2—Technical Specifications, 2021);

- **Soundbrenner Pulse** (Figure 1B), a wearable haptic metronome marketed primarily to musicians (Soundbrenner, 2026); and
- Peterson BodyBeat Pulse Solo, a compact clip-on device that converts metronome audio into tactile pulses (Peterson Electro-Musical Products, Inc., 2026).

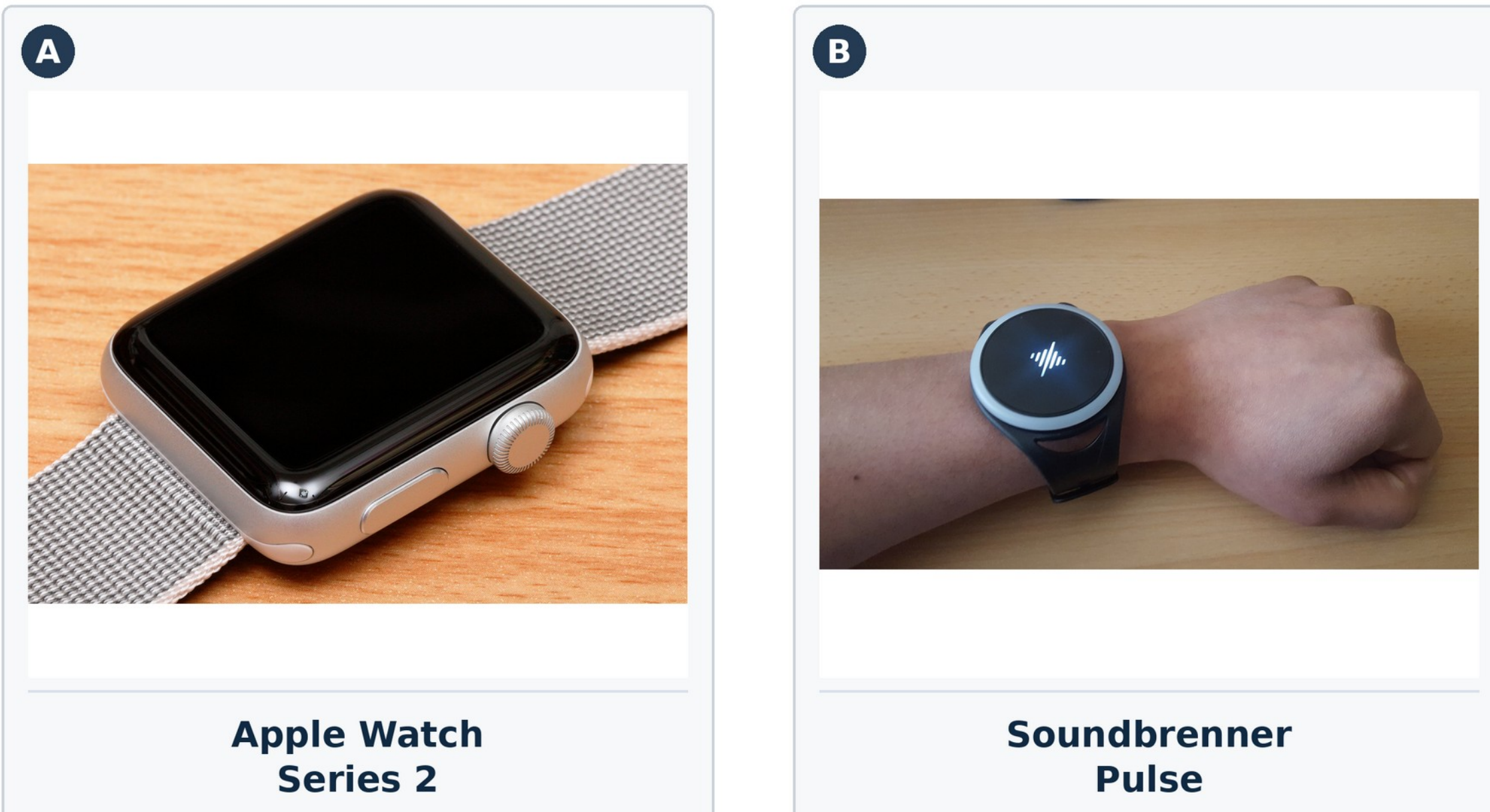

*Figure 1: Two of the wearable devices used as technology probes. (A) Illustrative photograph of an Apple Watch Series 2 (MiNe, 2017); and (B) Soundbrenner Pulse wearable haptic metronome. The Apple Watch used in the studies ran a bespoke haptic metronome application, which is not shown in panel A. The devices were used to expose user, device, and implementation requirements rather than evaluated as finished rehabilitation products. Image credits: (A) "MiNe-M 100-4006U (32298183471)" by MiNe, via Wikimedia Commons, CC BY 2.0 (https://creativecommons.org/licenses/by/2.0/), resized from the original; (B) photograph by the first author.*

Figure 1 illustrates the Apple Watch Series 2 and Soundbrenner Pulse. Together with the Peterson BodyBeat Pulse Solo, these systems provided contrasting haptic outputs, form factors, attachment methods, and interaction models through which rehabilitation-use requirements could be explored.

We selected and adapted devices pragmatically according to the requirements of each study and participant. In the formative sessions, we used the Apple Watch to test whether a familiar smartwatch form factor and a bespoke haptic metronome app could support cueing. This probe exposed issues around haptic intensity, screen size, app access, and interaction for participants with reduced dexterity or impaired upper limbs.

We used the Soundbrenner Pulse in several studies because it produced stronger vibration than the Apple Watch and allowed rhythm configuration through a companion app. This made it useful as a probe for haptic perceptibility, rhythm adjustment, and body placement. At the same time, the device and its interaction model exposed barriers around tapping, double tapping to start or stop the rhythm, dial use, device state, phone dependence, and use by people with impaired dexterity or cognitive burden.

We used the Peterson BodyBeat Pulse Solo in one community study as an alternative to the Soundbrenner Pulse. Its clip-on form factor allowed placement on clothing or close to the skin. The Pulse Solo receives an audio beat from an external metronome or compatible phone application and converts it into tactile pulses; it does not generate a rhythm independently. Product photographs are available on the manufacturer’s website (Peterson Electro-Musical Products, Inc., 2026). This probe helped expose requirements around equipment burden, attachment, carer support, and everyday use.

We did not interpret the limitations of these devices as product evaluations. Instead, we treated the differences between them as design evidence, revealing key requirements for future systems. Variations in haptic intensity highlighted the need for perceptible feedback, while differences in form factor informed requirements for device placement and fastening. Differences in interaction design, meanwhile, underscored the importance of low-burden controls. Collectively, these observations helped map the design space for future rehabilitation-oriented wearable haptic cueing systems.

## Use of gait and activity data

Gait and physical activity data were collected where feasible and appropriate. Variations in participants' mobility, health, transport access, and ability to attend laboratory appointments meant that equivalent quantitative data could not be obtained across all studies presented in this paper. Where possible, community participants completed assessments in an optical motion-capture gait laboratory. When laboratory assessment was impractical, alternative measures were used, including wearable sensors, mobile phone activity data, Fitbit summaries, and walking notes (Fitbit Official Site for Activity Trackers & More, 2020).

Because quantitative measures were not equivalent across studies, we used gait and activity data as contextual evidence rather than pooling outcomes. These data helped interpret the relationship between fixed cue tempo and variable walking cadence, day-to-day walking routines, differences between gait-laboratory and community cadence, and perceived benefit in relation to observed walking behaviour.

This approach was consistent with the pragmatic design orientation of the research. We used the most appropriate and feasible data sources in each setting rather than imposing one uniform measurement protocol across participants whose needs and contexts differed substantially.

## Cross-study analysis

We analysed the data across studies to identify recurring user, device, and implementation requirements. The analysis proceeded in four stages.

First, we collated study materials and extracted observations related to haptic perception, body placement, fastening, interaction, rhythm configuration, walking context, support needs, and perceived value. Second, we entered these observations into a cross-study matrix. The matrix compared findings by study, participant context, device, data source, and setting. Third, we grouped recurring observations into candidate themes using inductive thematic analysis (Braun & Clarke, 2006) informed by cross-study comparison. Fourth, we refined the themes into user requirement areas.

We retained a theme when it met at least one of two conditions. It either appeared across more than one study or data source, or it exposed an important boundary condition for future design. For example, repeated difficulties with haptic perception supported perceptibility as a core requirement. In contrast, some residential-care cases exposed boundary conditions where low residual sensation, cognitive burden, or safety concerns made wearable haptic cueing difficult or unsuitable without substantial support.

The analysis revealed seven recurring requirement areas:

1. haptic perceptibility and placement;
2. wearability and fastening;
3. interaction burden;
4. rhythm personalisation;
5. real-world walking context;
6. supported use by carers and clinicians; and

7. perceived value beyond gait outcomes.

We then organised these requirements into three interpretive categories. Essential design requirements were those that, when unmet, made the haptic devices difficult or impossible to use across multiple studies. Implementation requirements shaped feasibility, reduced burden, or reduced the risk of abandonment. Evaluation implications identified outcomes or forms of evidence that future studies should consider.

## Rigour and reflexivity

Variation across the six studies allowed us to examine requirements under different conditions of use. We used an explicit cross-study matrix to compare evidence across settings, devices, participant needs, and data collection methods, while recognising that each study contributed different evidence to the requirement areas.

We also retained negative and boundary cases. This was important as some of the most useful design insights came from cases where wearable haptic cueing was difficult, burdensome, or unsuitable. We required each final requirement area to be traceable to case-level evidence, such as participant quotations, observation notes, diaries, physiotherapist records, clinician comments, or gait and activity summaries.

A member of the research team independently reviewed the cross-study matrix, theme definitions, and representative evidence extracts for each requirement area. This review helped refine the distinction between essential design requirements, implementation requirements, and evaluation implications. It also helped check that the final requirements were supported by the available evidence.

Participants were not asked to review or validate the final cross-study themes following analysis. Consequently, the resulting requirement areas represent the authors' interpretation of the combined study data rather than a participant-validated framework. We acknowledge this as a limitation.

Direct quotations are identified by study and source where verbatim material was available. Evidence labelled as *interview* or *focus group* refers to verbatim transcript data, whereas evidence labelled as *interview summary*, *diary summary*, *observation note*, *physiotherapist record*, or *clinician comment* refers to contemporaneous records or detailed summaries produced during study analysis.

## Ethics and consent

The included studies received approval from The Open University Human Research Ethics Committee. The formative user sessions were covered by references HREC/2017/2633/Holland/1 and HREC/3000/Islam; the residential-care and community studies were covered by HREC/3000/Islam. The research was conducted in accordance with the Declaration of Helsinki. Participants gave informed consent for participation and anonymised reporting of findings. Pseudonyms are used for individual participants.

Risk assessments were undertaken as part of the research programme. Walking activities were supported or supervised where required by participants' mobility, balance, sensory, or cognitive needs. The off-the-shelf devices were used investigationally as technology probes rather than supplied as established rehabilitation treatments.

# Results

## Overview of cross-study findings

Across the six studies, the repurposed wearable devices acted as technology probes during conversations with participants that exposed three overarching requirement areas for future wearable haptic cueing systems.

First, users needed basic access to the cue. They had to perceive the vibration, wear the device at a suitable body location, and fasten and operate it either independently or with appropriate support, despite possible sensory, dexterity, or cognitive limitations.

Second, cueing had to accommodate variable walking rhythms and real-world walking contexts. Fixed rhythmic cues did not always align with changes in fatigue, terrain, interruptions, walking pace, or the differences between laboratory and community walking. At first glance, this appears to contrast with our earlier findings (Georgiou et al., 2015), in which stroke survivors did not favour manual cadence adjustment and emphasised the importance of simple controls for stopping cueing in demanding walking situations.

However, these findings are not necessarily inconsistent. They point to a common requirement: cueing must remain appropriate to the user's current context without imposing excessive cognitive burden. Participants in the earlier study did not favour explicit manual adjustment of cadence and considered automatic tempo adjustment desirable (Georgiou et al., 2015), whereas the studies considered here highlighted the limitations of fixed cueing across changing real-world conditions. Whether adaptability is best achieved through automatic context-aware adjustment, simple user controls such as pausing, or other interaction mechanisms remains an open question for future investigation.

Third, wearable haptic cueing often required support from physiotherapists, carers, family members, or researchers. In several cases, the cueing system did not function as an isolated user-device interaction, but as a supported rehabilitation technology.

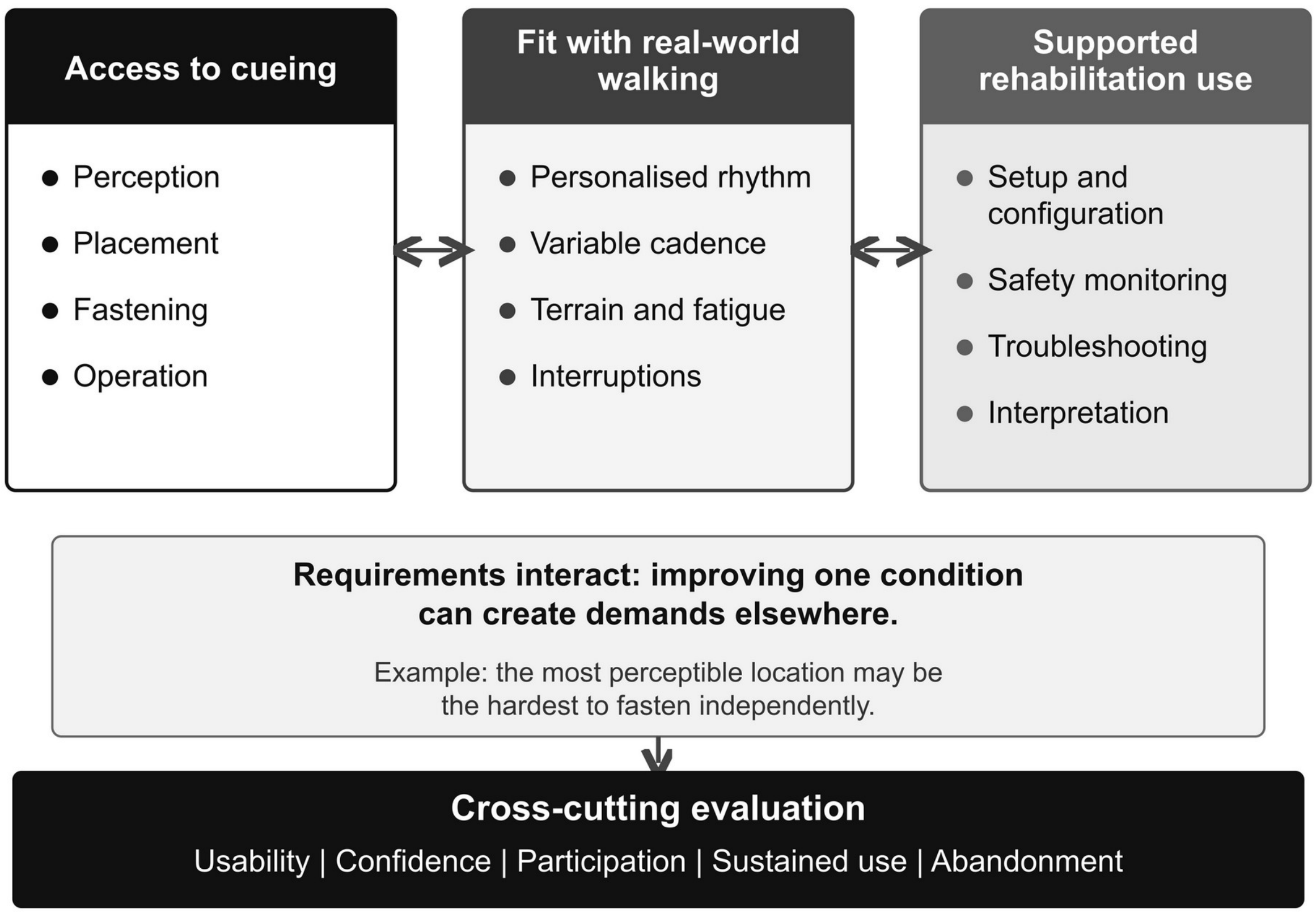

*Figure 2: Cross-study model of user requirements for wearable haptic cueing in neurological gait rehabilitation. Usable cueing depended on access to the cue, fit with real-world walking, and support for use. Bidirectional relationships indicate that requirements interacted and could create design trade-offs. Evaluation outcomes cut across all three areas.*

Figure 2 summarises the three connected requirement areas identified across the studies. The requirements are interdependent rather than sequential: addressing perception, placement, or interaction may create competing demands elsewhere in the system. Table 2 provides the supporting studies, evidence sources, and analytical role for each detailed requirement area.

*Table 2: Cross-study evidence supporting the seven recurring user-requirement areas.*

| Requirement area | Main supporting studies | Main evidence sources | Role in the analysis |
|---|---|---|---|
| Haptic perceptibility and placement | Studies 1, 2, 3, 5, 6 | Observation notes, interviews, physiotherapist records | Identified adjustable haptic output and flexible placement as entry requirements for cue perception |
| Wearability and fastening | Studies 1, 4, 5, 6 | Observation notes, interviews, diaries, carer-supported interview material | Showed that attachment must support one-handed or weak-handed use and repeated everyday wear |
| Interaction burden | Studies 1, 2, 4, 5, 6 | Observation notes, interviews, physiotherapist focus group, carer-supported interview material | Identified the need for simple controls, visible device state, and low cognitive and motor demand |
| Rhythm personalisation | Studies 1, 3, 4, 5, 6 | Interviews, diaries, activity summaries, gait analysis summaries | Showed that cue mode and tempo must be configurable across users, fatigue states, and walking contexts |
| Real-world walking context | Studies 2, 3, 4, 5, 6 | Interviews, diaries, physiotherapist records, activity summaries | Showed that terrain, interruptions, fatigue, supervision, and outdoor routes changed cueing feasibility |
| Supported use by carers and clinicians | Studies 2, 4, 6 | Physiotherapist focus group, carer-supported interview material, clinician comments | Showed that setup, safety monitoring, troubleshooting, and interpretation often depended on secondary users |
| Perceived value beyond gait outcomes | Studies 2, 3, 4, 5, 6 | Interviews, diaries, physiotherapist records, gait summaries | Identified confidence, routine formation, willingness to walk, and boundary cases as important evaluation considerations |

## Finding 1: Access to cueing depended on perception, placement, fastening, and operation

### *Haptic perceptibility and placement*

Haptic perceptibility varied substantially across users, body sides, and body locations. In the formative study, several participants with stroke or brain injury perceived the Apple Watch vibration poorly, and one participant could not perceive it at all at maximum strength (Study 1 observation note). The stronger haptic actuator in the Soundbrenner Pulse helped expose the importance of vibration intensity. However, stronger vibration alone did not solve the problem. Several participants still preferred to wear the device on the stronger arm because reduced sensation on the affected side made the rhythm harder to detect (Study 1 interview and observation notes).

This created a second design problem. The body location that supported perception was not always the easiest location for independent use. In Study 1, some participants could feel the cue better on the stronger side, but then had to use the weaker or affected arm to fasten and operate the device. This showed that haptic perceptibility and attachment cannot be designed separately. Future systems need to support both reliable cue perception and easy attachment at the perceptually useful body location. Strap designs such as Velcro, elasticated, magnetic, or slap-on mechanisms may reduce this burden. This trade-off is central: the best perceptual site may not be the easiest site for independent fastening or control.

The community studies reinforced this pattern. Hannah (pseudonym) reported that she could “feel the rhythm more strongly” on her stronger side and suggested that a cue placed “closer to [her] heart” might work better, possibly through a pendant-like form (Study 3 interview). She also described the cueing as helping her “march, march” and said that indoors she could “place her feet flat on the ground for the first time in many years” (Study 3 interview). Bob (pseudonym) preferred stronger vibration than Tracy (pseudonym), which showed that even community users with relatively stable self-management capacity had different perceptual needs (Study 5 and Study 6 interview summaries).

The residential-care study exposed an important boundary condition. Several residents could not perceive the vibration reliably even when physiotherapists increased the intensity or tried different placements on the arm, thigh, or calf (Study 2 physiotherapist records and focus-group summary). In these cases, wearable haptic cueing did not simply become less convenient. It became an unreliable cueing channel.

These probe findings show that haptic perceptibility is an entry requirement for wearable haptic cueing. Future systems need adjustable intensity, alternative pulse profiles, and flexible placement because neurological users may differ in residual sensation across body sides and body locations.

*Wearability and fastening*

Wearability problems were frequent and practical. In Study 1, conventional watch-style straps created difficulty for participants with hemiparesis, especially when the device had to be worn on the stronger side but fastened using the weaker or affected arm (Study 1 observation note). The formative sessions showed that familiar wearable form factors can still become difficult when users have upper-limb asymmetry, reduced dexterity, or reduced tactile feedback.

Study 4 and Study 6 developed this requirement further. Rosie (pseudonym) struggled with watch-like attachment and often relied on help from her mother or friend (Study 4 interview summary). Her mother suggested that a more suitable solution would use Velcro or elasticated fastening rather than small watch-like fixtures. She also suggested that the controls should be easier to access (Study 4 carer-supported interview material). Bob, who had reduced strength and dexterity in both arms, requested a customised strap. A Velcro-based strap improved wearability, but the bond between the device and strap loosened with everyday use (Study 6 diary summary). This exposed a second requirement: fastening must be easy to use and mechanically durable.

Wearability also included social acceptability. Tracy rejected the idea of a bulky ankle-worn device because it looked, in her words, as if she had "recently been released from prison and was being monitored" (Study 5 interview). Bob preferred a watch-like appearance and asked why the device could not also show the time, suggesting that multifunctionality may reduce the burden of wearing an additional device (Study 6 interview summary).

These findings show that attachment and form factor are central design requirements. A device may generate an appropriate rhythm, but cueing cannot become part of walking practice if users cannot attach it independently, wear it comfortably at the perceptually useful location, or rely on the fastening during everyday movement.

*Interaction burden*

Interaction burden also limited access to cueing. In Study 1, the smartwatch probe exposed interaction barriers around screen size and app launching. Some participants found the Apple Watch screen too small and struggled to locate or launch the bespoke haptic metronome application (Study 1 interview and observation notes). The Soundbrenner Pulse exposed a different interaction profile. Users could often perceive the vibration more easily, but the device still required taps, double taps to start or stop the rhythm, dial turns, and interpretation of device state. In the formative study, only one participant with relatively better physical and sensory function could reliably perform all required interactions independently (Study 1 observation note).

In the residential-care study, physiotherapists reported a related but clinically distinct issue. Some residents could physically attempt the task, but could not maintain attention to both walking and the cue. One resident completed the physical trial but could not continue because the combined attentional burden was too high (Study 2 physiotherapist focus-group summary). In this setting, interaction burden merged with cognitive load and safety.

The community studies produced similar evidence. Tracy reported that the double-tap command did not always work and that she sometimes could not stop the rhythm when she stopped walking (Study 5 interview). Bob wanted the interaction model reduced to a small number of reliable actions: on/off, rhythm adjustment, and a visible beats-per-minute display (Study 6 interview summary). Rosie partly depended on others to operate the device, and her mother suggested larger, more straightforward buttons (Study 4 carer-supported interview material).

These findings show that future rehabilitation-oriented wearable haptic cueing systems should not inherit musician-oriented or smartwatch-style interaction assumptions without redesign. Users need controls that are explicit, reliable, visible, and usable before, during, and after walking.

## Finding 2: Cueing had to fit variable walking rhythms and real-world walking contexts

### *Rhythm personalisation*

No single rhythm strategy appeared sufficient across the studies. In this paper, **bipedal cueing** refers to cues intended to support both left and right steps, whereas **monopedal cueing** refers to cueing one side or one step within the gait cycle. In Study 1, some participants preferred bipedal cueing, while others with slower gait or higher cognitive demand found monopedal cueing easier to follow (Study 1 interview summaries). One participant with a very slow walking pace could tap along when stationary but could not synchronise stepping to the imposed bipedal rhythm (Study 1 observation note). This showed that understanding a cue and synchronising gait to that cue were different demands.

The community studies showed that rhythm requirements changed by person, day, and context. Hannah adjusted the rhythm according to fatigue and daily variation (Study 3 diary and interview). Rosie kept a more stable rhythm arrangement during the study period (Study 4 diary summary). Bob preferred to start with a slower cadence and then increase the tempo once he had established his pace (Study 6 interview summary). Tracy exposed a different limit: her hilly walking routes caused the fixed cue to feel “too slow” at some points and “too fast” at others (Study 5 interview).

Bob’s repeat study after one year added an important observation. He completed two periods of wearable haptic cueing use separated by a one-year gap. Gait analysis data indicated that he could reproduce a similar walking rhythm across the two study periods (Study 6 gait analysis summary). This observation is relevant because rhythm following and entrainment are central mechanisms in rhythmic cueing research (Thaut et al., 1997; Hurt et al., 1998; Clayton et al., 2005; Grahn & Brett, 2007; Hove et al., 2012). However, we do not treat this as evidence of clinical efficacy. Instead, we treat it as a design-relevant observation: some users may be able to re-establish a preferred rhythm over time, so future systems should support rhythm recall, rhythm adjustment, and clinician review of rhythm settings.

Bob’s activity summaries also showed that cadence in everyday walking was lower than cadence in gait-laboratory conditions, likely because of fatigue, crossings, obstacles, and route interruptions (Study 6 activity summary). This reinforced the same requirement from a different angle. A rhythm that seems suitable in controlled assessment may not hold in community walking.

These findings show that rhythm personalisation shaped feasibility throughout the rehabilitation pathway. Future systems should support personalised tempo, monopedal and bipedal cueing options, and simple rhythm adjustment. The findings also point toward adaptive cueing as a future design direction, but the present studies did not test adaptive cueing directly.

### *Fit with real-world walking contexts*

The practical meaning of wearable haptic cueing changed across settings. Because we used the devices as probes rather than controlled interventions, these contextual differences were analytically useful. They showed how cueing requirements changed between supervised residential care, indoor community walking, outdoor routes, hills, interruptions, and accompanied walking.

In Study 2, physiotherapists used the device in a supervised residential-care setting. Even with physiotherapist support, many residents could not use the cueing effectively because low residual sensation, low mobility, balance concerns, or cognitive burden left too little capacity for safe cue following (Study 2 physiotherapist records and focus-group summary). This was not simply a smaller benefit. For some residents, wearable haptic cueing was unsuitable without substantial support or redesign.

In the community studies, users faced different contextual barriers. Hannah found the device easier to use indoors in long corridors than outdoors, where conversation, surrounding activity, and safety concerns competed for attention (Study 3 interview). Rosie mainly used the device outdoors and often walked with her mother or friend (Study 4 diary summary). Tracy's hilly routes made fixed-tempo cueing hard to sustain and contributed to foot discomfort when she walked more than usual (Study 5 interview and physiotherapist summary). Bob's diary notes described occasions where the device fell from the strap during everyday activity, including while greeting another person (Study 6 diary summary).

Across the studies, corridor walking and laboratory assessment did not expose the full burden of everyday use. Real walking involved terrain, route variation, fatigue, distractions, social interaction, and interruptions. Future wearable haptic cueing systems therefore need to work in real-world walking contexts, not only in controlled walking tasks.

## Finding 3: Cueing functioned as a supported rehabilitation technology

### *Supported use by carers and clinicians*

Wearable haptic cueing often required support from people around the user. In Study 2, physiotherapists selected residents, configured the device, varied rhythm and intensity, trialled different placements, and judged safety. Their reflections also moved beyond immediate usability. They asked for automatic step counting or other outputs that could reduce workload and improve interpretability (Study 2 physiotherapist focus-group summary).

In community use, family members and clinicians also shaped cueing use. Rosie's mother and friend helped with wearing, operating, and interpreting the device (Study 4 interview summary). Her mother suggested improvements to both the hardware and the study materials, including larger buttons and richer walking logs that captured surfaces and turning demands (Study 4 carer-supported interview material). Bob lived independently, but physiotherapists still helped interpret changes in his walking pattern and consider what objective data might mean clinically (Study 6 clinician comments).

Even apparently self-managed cases contained hidden support structures. Researcher setup, training, charging support, and follow-up contact reduced the burden of use. Future clinical or community provision would need to reproduce some of this support if users cannot manage the device independently.

These findings show that wearable haptic cueing often functions as a supported rehabilitation technology. Future systems need workflows for setup, charging, troubleshooting, safety monitoring, rhythm adjustment, and interpretation by carers or clinicians where needed.

### *Perceived value beyond gait outcomes*

Several community participants described value beyond narrowly defined gait outcomes. Hannah reported increased confidence and more willingness to walk (Study 3 interview). Rosie associated the device with longer walks, more confidence, and better mood (Study 4 interview summary). Bob described the cueing as helping him establish a walking routine and maintain a steadier pace (Study 6 interview and diary summary). These accounts came mainly from interviews and diaries, so we interpret them as perceived value rather than evidence of therapeutic effect.

Negative and boundary cases also shaped this finding. Tracy walked more during the study, but this did not translate into clear gait improvement and contributed to foot discomfort (Study 5 physiotherapist summary and gait summary). In the residential-care study, many residents could not use the intervention effectively, so perceived value could not be assumed.

One residential-care case was particularly notable. A physiotherapist record described a resident with functional neurological disorder completing a supervised 10-metre walk during cued use after a prolonged period (two years) without walking (Study 2 physiotherapist record). Functional neurological disorder can involve genuine and disabling motor symptoms, including gait impairment, without structural neurological damage that fully explains the presentation (Bennett et al., 2021). We present this observation cautiously. It does not show that wearable haptic cueing treats functional neurological disorder. However, it suggests that supervised haptic cueing may affect task engagement, confidence, or willingness to attempt walking in some cases, which should be explored without treating this observation as efficacy evidence.

Taken together, these findings show that future studies should not assess wearable haptic cueing only through spatiotemporal gait parameters. They should also consider whether users feel more confident, walk more willingly, form routines, or need support to sustain use. These outcomes should complement gait measures rather than replace them.

## User, device, and implementation requirements

Wearable haptic cueing must be designed around the conditions required for real rehabilitation use: perceiving the cue, wearing the device, operating it safely, configuring the rhythm, using it in real walking contexts, and supporting use through carers or clinicians where needed.

These requirements are not product-specific evaluations of the probe devices. They are design lessons derived from how the probes succeeded, failed, or required support across rehabilitation contexts.

Table 3 summarises the user, device, and implementation requirements derived from the cross-study findings.

*Table 3: User, device, and implementation requirements for future rehabilitation-oriented wearable haptic cueing systems.*

| Area | Requirement | Why it matters for gait rehabilitation | Design implication |
|---|---|---|---|
| Haptic perceptibility | Adjustable haptic intensity and, where possible, pulse profile | If the rhythm cannot be perceived reliably, rhythm following cannot occur | Support configurable vibration strength and alternative pulse settings |
| Placement | Flexible body placement | Affected limbs may be poor cue sites; stronger sides may be easier to perceive but harder to fasten | Provide modular attachment options for wrist, upper arm, torso, thigh, or other suitable body sites |
| Fastening | One-handed or weak-handed attachment with everyday robustness | Cueing fails if the device cannot be attached at the site needed for reliable perception | Use Velcro, elasticated, magnetic, or clinician-configurable attachment systems that withstand repeated use |
| Interaction | Low-burden controls and visible state | Unreliable controls disrupt safe cue use during walking | Use a small set of explicit controls, visible rhythm state, and readable beats per minute (BPM) or status |
| Rhythm configuration | Personalised cue mode and tempo | Fixed cue structures can conflict with variable real-world cadence | Support user or clinician configuration of cue mode and tempo with minimal setup burden |
| Real-world walking context | Robustness to terrain, interruptions, fatigue, and social activity | Terrain and interruptions can disrupt cue-step synchronisation | Evaluate and configure systems for actual walking contexts rather than clinic corridors alone |
| Support from carers and clinicians | Secondary-user workflows | Cueing often depends on shared setup, safety monitoring, troubleshooting, and interpretation | Design setup, charging, troubleshooting, and interpretation pathways for shared use |

| Area | Requirement | Why it matters for gait rehabilitation | Design implication |
|---|---|---|---|
| Evaluation | Outcomes beyond gait mechanics | Cueing may alter confidence and engagement even when gait change is limited or mixed | Include usability, confidence, participation, sustained use, and abandonment alongside gait measures |

Overall, the results show that future wearable haptic cueing systems need to be designed around access, context, and support. Access concerns whether users can perceive, wear, fasten, and operate the cueing system. Context concerns whether the rhythm can fit real walking conditions. Support concerns whether carers, clinicians, or researchers need to help with setup, safety, interpretation, or sustained use. These findings form the basis for the discussion, where we consider implications for rehabilitation technology design, clinical implementation, and future evaluation.

# Discussion

## Principal findings

We found that wearable haptic cueing does not become useful in neurological gait rehabilitation simply because a device can generate a rhythmic vibration. As summarised in Figure 2, use depended on three connected conditions: access to the cue, fit with real-world walking, and support for use.

Access to the cue required reliable haptic perception, suitable body placement, manageable fastening, and low-burden controls. Fit with real-world walking required rhythm personalisation and robustness to fatigue, terrain, interruptions, and changing cadence. Support for use involved carers, clinicians, physiotherapists, family members, or researchers who helped with setup, safety, troubleshooting, interpretation, and sustained use.

These findings shift the focus from whether wearable haptic cueing can deliver a rhythm to what users need so that cueing can become part of rehabilitation practice.

## Technology probes and the design space for wearable haptic cueing

We did not evaluate the Apple Watch, Soundbrenner Pulse, or Peterson BodyBeat Pulse Solo as finished rehabilitation products. We used them as technology probes to make the design space visible. Their strengths and limitations helped us identify what future wearable haptic cueing systems need to support: reliable cue perception, flexible placement, accessible attachment, low-burden interaction, rhythm personalisation, real-world walking use, and supported use by carers or clinicians.

This contribution differs from much of the wearable haptic cueing literature. Existing reviews and trials have mainly asked whether cueing can improve gait, balance, freezing of gait, or motor outcomes under defined conditions (Sweeney et al., 2019; De Angelis et al., 2021; Gonçalves et al., 2021; Zhang et al., 2024). Those questions are important, but they come after a prior design question: what must be true for a person with neurological gait impairment to perceive, wear, operate, configure, and sustain wearable haptic cueing in the first place?

This paper builds on a programme of prior work exploring wearable haptic cueing for gait rehabilitation across stroke, brain injury, and Huntington's disease. Earlier studies and case-based deployments showed that body-worn haptic rhythms could be piloted in clinical and community contexts, but they primarily reported individual prototypes, cases, or early feasibility findings (Georgiou et al., 2020a; Georgiou et al., 2020b; Islam et al., 2018; Islam, 2023). What remains underdeveloped is a consolidated account of the user requirements that emerge across settings, especially when wearable haptic cueing moves beyond controlled or single-case use into residential care and community self-management. The present paper addresses that gap by synthesising cross-setting evidence on perception, placement, fastening, interaction, rhythm configuration, real-world walking, and supported use.

Our findings show why this prior design question matters. A cueing system may show short-term gait effects in a supervised setting but still fail as a rehabilitation technology if users cannot feel the cue, fasten the device, interpret its state, adjust the rhythm, or use it safely in everyday walking. This aligns with wider rehabilitation technology literature, which shows that adoption depends not only on technical performance, but also on contextual fit, stakeholder involvement, service capacity, support burden, and user workload (Mitchell et al., 2023; Cheng et al., 2024; MacNeil et al., 2024; Layton et al., 2024).

The specifically haptic and gait-related contribution is that cue perception is the entry condition for rhythm following. Attachment determines whether users can wear the cue at a perceptually useful body location. Fixed-tempo cueing becomes fragile when walking cadence changes because of fatigue, terrain, interruptions, or outdoor routes. These are not minor usability details. They determine whether wearable haptic cueing can be used at all (Demain et al., 2013; Casamassima et al., 2014; Chen et al., 2017; Heldman et al., 2016).

## Implications for rehabilitation technology design

The findings point to several design requirements for future wearable haptic cueing systems.

First, systems should provide adjustable haptic output. A single vibration intensity did not work across users, body sides, or body locations. Some participants could not perceive weaker vibration, especially on an affected limb. Others needed stronger output during walking, where movement and attention may make haptic cues harder to detect. Future systems should therefore support adjustable intensity and, where possible, alternative pulse profiles.

Second, systems should support flexible placement. The wrist was not always the best location. Some users perceived the cue better on the stronger side, while others suggested different body locations, such as closer to the torso. Designers should not assume that the clinically affected limb is the best cueing site. Future systems should support modular placement on the wrist, upper arm, torso, thigh, calf, clothing, or other appropriate locations.

Third, systems should use accessible fastening. Fastening is not a small ergonomic issue in neurological rehabilitation. Users with hemiparesis, reduced upper-limb strength, or impaired dexterity may struggle to fasten a device at the very location where they can best perceive the cue. Future systems should support one-handed or weak-handed attachment through mechanisms such as Velcro, elasticated straps, magnetic closures, slap-on forms, clips, or attachments that clinicians or carers can configure. These mechanisms also need to withstand repeated everyday use.

Fourth, systems should reduce interaction burden. Users should not need to manage small screens, hidden device states, unreliable gestures, or complex tapping sequences while preparing to walk or while walking. Future systems should provide a small number of explicit controls, clear on/off state, visible rhythm settings, and simple rhythm adjustment. Where possible, designers should separate clinical configuration from everyday user operation.

Fifth, systems should support rhythm personalisation. Users differed in preferred tempo, cueing mode, and ability to follow monopedal or bipedal rhythms. Some users needed slower rhythms to establish walking; others needed adjustment across days or routes.

Future systems should allow users or clinicians to personalise rhythm settings with minimal burden. This may include saving preferred tempos, supporting slower starting rhythms, enabling simple adjustment across days or routes, and offering different cueing modes such as monopedal or bipedal cueing. In our studies, fixed-tempo cueing became difficult when walking conditions changed, including during hills, fatigue, crossings, obstacles, or shifts in cadence. Future systems may therefore need some form of rhythm flexibility, whether through simple user adjustment, clinician-mediated configuration, or automated adaptation. However, more flexible or adaptive cueing may also introduce new demands around interaction, interpretation, and safety. We therefore present rhythm adaptation as an important design direction, but not as a solution demonstrated by the present studies.

Figure 2 also highlights that these requirements should not be addressed independently. Increasing haptic intensity, changing body placement, simplifying controls, or introducing carer support may solve one problem while creating another. Future systems therefore require configuration at system level rather than isolated optimisation of individual features.

## Implications for clinical implementation and community rehabilitation

Wearable haptic cueing also has implications for rehabilitation delivery. Many people with long-term neurological conditions live in the community and do not receive the same level of regular support available in hospital or residential-care settings. Community rehabilitation can be intermittent and resource-constrained, and home-based rehabilitation technologies often need to bridge support gaps between formal contacts (Ayoade et al., 2013; Chen et al., 2019). Access may depend on referral pathways, local service capacity, geography, cost, and availability of specialist neurophysiotherapy. Global rehabilitation need is also increasing, and rehabilitation services need to move closer to communities to reach more people who need support (Cieza et al., 2020).

This matters for wearable haptic cueing because community users may need ongoing support, not only a device. In our studies, carers, clinicians, physiotherapists, family members, and researchers often helped with setup, troubleshooting, safety monitoring, rhythm configuration, charging, and interpretation. Future service models should therefore treat wearable haptic cueing as a supported rehabilitation technology rather than a device handed to an isolated user (Ayoade et al., 2013; Keogh et al., 2021; Horsky et al., 2012).

Wearable systems may also help address gaps between formal rehabilitation appointments. Wearable movement sensors already show promise for tracking movement, activity, gait, and rehabilitation progress in real-world settings (Iosa et al., 2016; Díaz et al., 2019; Porciuncula et al., 2018; Rodgers et al., 2019; Angelini et al., 2020; Auepanwiriyakul et al., 2020; Keogh et al., 2021). If future wearable haptic cueing systems combine cue delivery with sensing, they may support both intervention and monitoring. For example, they could help clinicians understand whether people use cueing regularly, whether walking routines change over time, whether cadence or activity declines, and whether users need review or adjustment.

This may be especially important for degenerative neurological conditions such as Parkinson's disease and Huntington's disease. Wearable sensors are increasingly studied for monitoring gait, tremor, bradykinesia, dyskinesia, and other movement features in Parkinson's disease (Bougea, 2025). Reviews and recent digital studies also show growing interest in wearable and portable sensors for Huntington's disease, although clinical utility and standardisation remain ongoing challenges (Tortelli et al., 2021; Nunes et al., 2024). These developments suggest a future direction where wearable haptic cueing systems do more than deliver prompts. They may also contribute data that helps clinicians detect change, review rehabilitation adherence, personalise treatment, and identify decline earlier.

However, this future depends on design. Users must be able to wear the system, trust it, understand it, and integrate it into daily life. Clinicians must receive data that is interpretable and useful rather than burdensome. Carers must know how to support use without becoming overburdened. Healthcare systems must decide how such data connects to existing clinical workflows. Our findings therefore provide early requirements for this broader vision: future systems need accessible hardware, low-burden interaction, personalised cueing, clear support workflows, and clinically meaningful data pathways.

## Suitability, configuration, and non-suitability

Wearable haptic cueing should not be introduced as a generic intervention for all neurological gait impairment. Our residential-care study showed this clearly. Some residents had too little residual haptic perception, too much cognitive burden, too little balance reserve, or too much support need for current forms of wearable haptic cueing to be useful or safe.

This finding challenges an overly simple view of inclusive design. Inclusive design does not mean one device should work for everyone. It means designers and clinicians must understand the conditions needed for safe and meaningful use. In some cases, a user may need stronger haptics, a different body location, carer support, or supervised setup. In other cases, wearable haptic cueing may not be appropriate.

Table 4 summarises suitability domains suggested by the findings. We present this table as a design and service-planning aid, not as a validated clinical screening tool.

*Table 4: Suggested suitability domains for configuring wearable haptic cueing in rehabilitation practice.*

| Screening domain | Why it matters | Example question |
|---|---|---|
| Haptic perception | The user must perceive rhythm reliably for rhythm following to be possible | Can the user detect the cue at intended body sites? |
| Side-specific sensation | The affected limb may be the least reliable perceptual site | Which body side or location gives reliable perception? |
| Dexterity | The user must be able to attach and control the device | Can the user put it on, take it off, and start or stop it independently or with support? |
| Cognitive load | The user must attend to the cue while walking without unsafe distraction | Can the user follow the rhythm without losing track of the walking task? |
| Balance and falls risk | Cueing may divide attention or increase risk for users with complex support needs | Is supervised use required during setup or walking? |
| Walking context | Terrain and route features affect rhythm matching | Will use occur indoors, outdoors, on hills, or around crossings and interruptions? |
| Support arrangements | Setup and review may depend on carers or clinicians | Who configures, charges, troubleshoots, and reviews use? |

## Implications for evaluation

Future studies should not evaluate wearable haptic cueing only through controlled gait assessments. In our studies, corridor walking and gait-laboratory findings did not fully capture community use. Users encountered hills, fatigue, interruptions, conversation, social activity, safety concerns, and device-management problems. These factors shaped whether cueing was feasible. This is consistent with broader wearable-sensor work showing that gait and mobility measured in everyday life can differ meaningfully from gait observed in laboratory or clinic settings (Brodie et al., 2016; Angelini et al., 2020).

Future evaluations should therefore include real-world walking conditions. They should assess whether users can wear the device, operate it, adjust rhythm, sustain use, and manage the cue in everyday routines. Gait parameters remain important, but they should sit alongside usability, perceived confidence, support burden, and sustained use.

Our findings also suggest that perceived value can extend beyond gait mechanics. Some participants described confidence, routine formation, willingness to walk, and mood-related benefits. We do not treat these accounts as evidence of therapeutic efficacy. Instead, we treat them as important evaluation targets. If wearable haptic cueing helps a person walk more confidently or establish a walking routine, future studies should be able to capture that value.

## Limitations

This study has several limitations.

First, the sample was small and heterogeneous. The findings should therefore be interpreted as transferable design propositions rather than generalisable clinical conclusions.

Second, the studies differed in setting, duration, device, and data sources. This variation was deliberate and supported the technology-probe approach, but it also limited direct comparison across studies.

Third, the data corpus was uneven. Some studies had richer interview, diary, gait, or activity data than others. We addressed this through the cross-study matrix, but the strength of evidence still varied by requirement area.

Fourth, the gait and activity data were selective and contextual. We did not collect uniform gait measures across all participants, and we did not pool quantitative outcomes. The paper therefore cannot make claims about treatment efficacy.

Fifth, the analysis was interpretive. A member of the research team reviewed the matrix, theme definitions, and representative evidence extracts, but we did not independently double-code the full dataset. We also did not ask participants to review the final cross-study themes.

Sixth, the studies were short. Except for Bob's repeat study after one year, we had limited evidence about long-term adoption, abandonment, maintenance, or sustained benefit.

Seventh, the devices are no longer necessarily current or optimal rehabilitation products. Their value in this analysis lies in their role as probes. They exposed requirements around perception, placement, fastening, interaction, rhythm configuration, and support. The findings should therefore be read as requirements for future systems, not as product-specific conclusions about the Apple Watch, Soundbrenner Pulse, or Peterson BodyBeat Pulse Solo.

Eighth, the studies did not directly examine integration with healthcare systems. We argue that future wearable haptic cueing systems may support community rehabilitation, monitoring, and personalised care, but this remains a future research direction. Future studies should examine data governance, clinician workload, interoperability, clinical decision-making, and user trust.

Finally, transferability to people with more severe impairment may be limited, especially where profound sensory loss, cognitive burden, or safety concerns dominate. In such cases, wearable haptic cueing may require substantial redesign, supervised use, or may not be appropriate.

## Future work

Future work should not move directly to larger efficacy trials with the same device assumptions. A staged programme is needed.

First, researchers and designers should develop rehabilitation-oriented wearable haptic cueing systems that address the requirements identified here: adjustable haptics, flexible placement, accessible fastening, simple controls, rhythm personalisation, real-world robustness, and supported use.

Second, future studies should test these systems in everyday walking contexts, not only in gait laboratories or corridors. They should examine how users manage cueing during fatigue, outdoor walking, terrain changes, interruptions, and social activity.

Third, future systems should explore longer-term use. Longer studies could show whether people sustain cueing routines, abandon devices, require support, or change walking behaviour over time.

Fourth, future work should explore systems that combine cueing and sensing. Such systems could support rhythm delivery while also collecting information about walking routines, activity levels, cadence changes, adherence, and possible decline. This may be especially relevant for degenerative neurological conditions such as Parkinson’s disease and Huntington’s disease.

Finally, future research should examine how wearable cueing and sensing data can connect with rehabilitation services. If wearable systems produce clinically meaningful data, they could supplement outpatient physiotherapy records, support remote review, identify changing needs, and contribute to more personalised rehabilitation planning. This will require careful design around privacy, data interpretation, clinician workload, and integration with existing care pathways.

## Conclusion

Wearable haptic cueing for neurological gait rehabilitation should not be designed as cue delivery alone. Across six linked studies, repurposed wearable devices acted as technology probes that exposed the conditions required for real rehabilitation use: reliable cue perception, flexible placement, accessible fastening, low-burden operation, rhythm personalisation, robustness to real-world walking contexts, and support from carers or clinicians where needed.

These findings should not be read as product-specific evaluations of the devices used. They are user requirements for future rehabilitation-oriented wearable haptic cueing systems. The next generation of systems should therefore be designed not as wearable metronomes, but as rehabilitation technologies embedded within user capabilities, walking routines, support arrangements, and, where appropriate, longer-term care pathways.

## Acknowledgments

We thank the participants, their families and carers, and the physiotherapists who contributed to the studies.

## Funding

Riasat Islam was supported by the Goldcrest Charitable Trust.

## Declaration of interest

The authors report there are no competing interests to declare.

## Data availability statement

Anonymised quotations supporting the findings are included in this article. The underlying participant-level records, including interview transcripts and clinical information, are not publicly shared because participant consent permits sharing anonymised quotations rather than the underlying records.

## Declaration of generative AI use

OpenAI tools, including GPT-6.1 and earlier ChatGPT versions (5.5, 5.6, 6), were used to assist with copyediting, language refinement, reference lookup, and submission-file preparation.